\documentclass[epj]{svjour}

\usepackage{amsmath}
\usepackage{ulem}
\usepackage{graphicx}
\usepackage{graphics}

\newcommand{\grad}{\ensuremath{^\circ}}

\begin{document}
\title{A new type of temperature driven reorientation transition in magnetic thin films}

\author{F. K\"{o}rmann\inst{1} \and S. Schwieger\inst{2} \and J. Kienert\inst{1} \and W. Nolting\inst{1}
}                     

\institute{Lehrstuhl Festk{\"o}rpertheorie, Institut f{\"u}r Physik, Humboldt-Universit{\"a}t zu Berlin, 
  Newtonstr. 15, 12489 Berlin, Germany \and Technische Universit\"at Ilmenau, Theoretische Physik I,
           Postfach 10 05 65, 98684 Ilmenau, Germany}
\date{Received: date / Revised version: date}
	\abstract{We present a new type of temperature driven spin reorientation transition (SRT) in thin films. It can occur when the lattice and the shape anisotropy 
favor different easy directions of the magnetization. Due to different temperature dependencies of the two  contributions  the {\it effective} anisotropy 
may change its sign and thus the direction of the magnetization as a function of temperature may change. Contrary to the well-known reorientation transition caused by 
competing surface and bulk anisotropy contributions the reorientation that we discuss is also found in film systems with a uniform lattice anisotropy. 
The results of our theoretical model study may have experimental relevance for film systems with positive lattice anisotropy, as e.g. thin iron films grown on copper.
} 

\maketitle
\section{Introduction}
In the last years nanophysics has attracted considerable attention. With respect to technological application one objective is to create ultrahigh density magnetic data storage media. For the functionality of those components the magnetic anisotropies  in ultrathin film systems are of crucial importance.
The therewith directly connected reorientation transition of the magnetization of thin films and multilayer systems as a function of film thickness, temperature, applied magnetic field, and 
structural variations has been studied intensively during the last years. The orientation of the magnetization may vary between the perpendicular (out-of-plane)  and the parallel (in-plane) direction to 
the film plane, as well as between certain directions within the film plane. Concerning the first type of reorientation (in-plane vs. out-of-plane) most frequently the reorientation from the 
perpendicular to the in-plane magnetization with increasing film thickness and temperature has been observed, in particular for Fe/Cu(001) \cite{wandlitz1}, 
Fe/Ag(001) \cite{wandlitz2}, Fe/Cr(110) \cite{wandlitz3}, CeH$\rm_2$/Fe(111) \cite{wandlitz4},\\
Fe/Tb(111) \cite{wandlitz5}, Co/Au(111) \cite{cobalt1},
Co/Ru(111) \cite{cobalt2},\\
Co/Rh(111) \cite{cobalt3}, Co/Pd(111) \cite{cobalt4} and other film and multilayer systems.

The mechanism behind these temperature or thickness dependent reorientations can be described as a surface or interface effect. The lattice anisotropies 
at the film surface/interface and within the film interior have different signs, one favoring parallel and the other favoring perpendicular alignment of the 
magnetization to the film plane. In addition dipole-dipole interactions induce the so-called shape-anisotropy which favors an in-plane magnetization for all layers.
However, even if the layer resolved anisotropy alone favored different directions of the magnetization at the film surface and in the film interior, the much stronger ferromagnetic Heisenberg exchange coupling aligns the magnetizations of the different layers almost parallel. This holds especially for ultrathin films. For instance, in Ref. \cite{usadel} parallel alignment was found  
for a Fe-like film system up to 5 ML. The surface and interior anisotropies thus compete to determine the 
direction of the magnetization. For thinner films the surface contribution dominates while for thicker 
films $d>d_{crit}$ the film interior contribution overcomes the surface part. 

Competing surface and interior contributions can also lead to a temperature driven reorientation transition. 
The effective anisotropy energy, i.e. the 
sum of the lattice and shape anisotropy  energies, decreases and vanishes with the magnetization. On the other hand  with increasing $T$ the magnetization of the 
surface layer of a film is reduced more as compared to the film interior \cite{roland}. Thus  when at $T=0$ the surface contribution to the anisotropy dominates 
it may become smaller than the interior contribution at a certain critical temperature $T_{reo}$. This kind of temperature driven spin reorientation transition due 
to competing surface and bulk contributions is well understood today \cite{wandlitz17,wandlitz18,wandlitz19,wandlitz20,wandlitz21,wandlitz22,wandlitz24}.

 Let us now discuss translational invariant systems which have no surface and therefore no competing contributions as described above. In these systems, a competition between shape and lattice anisotropy is still possible if the lattice anisotropy favors out of plane orientation of the magnetization. Such a competition can lead to a temperature driven reorientation transition if the lattice and the shape anisotropy possessed different temperature dependencies. This transition can not occur if  the contributions due to  the second order lattice anisotropy $\tilde{K}_2(T)$ and shape anisotropy $\tilde{g}_0(T)$ have the same temperature dependence (see e.g. Refs. \cite{wandlitz,herrmann}). In this paper, we use a method resulting in different temperature dependencies for the lattice and shape anisotropy. As a consequence, a new type of temperature driven reorientation transition is possible, that is solely caused by the competition between lattice and shape anisotropy.

Before giving a more detailed account of our theoretical description in the next section we briefly summarize the main results for the effective anisotropy. The 
effective anisotropy  field is given by 
\begin{eqnarray}
K_{2eff}(T) &=& (2K_{2} C(T) -\frac{3}{2}Dg_{0})\langle S_{z\prime}\rangle(T)\label{Gl1}\\
 &=& \tilde{K_{2}}(T) + \tilde{g_0}(T)\label{g0effekt}
\end{eqnarray}
The sign of the effective anisotropy field determines the magnetic 
easy axis. The latter is aligned perpendicular to the film plane for positive effective anisotropy  fields $K_{2eff}>0$  and parallel to the film plane for
negative  effective anisotropy fields $K_{2eff}<0$. 
The effective anisotropy field must not be confused with the magnetic anisotropy energy since both quantities may show very different temperature dependencies. However, the anisotropy field discussed here directly determines the excitation energies and therewith also the magnetic free energy. Therefore the magnetic reorientation can be discussed solely in terms of this effective anisotropy field instead of considering the magnetic anisotropy energy explicitly.

In Eq. (\ref{Gl1}) $K_{2}$ denotes the microscopic anisotropy parameter and $g_0$ the dipolar coupling strength (note that both quantities are parameters  and that they are assumed to be temperature independent  in the following). $D$ is a constant determined by the lattice structure and $\langle S_{z\prime}(T) \rangle$ denotes the norm of the magnetization \footnote{Note that the norm of the magnetization $\langle S_{z\prime}(T) \rangle$ and its component aligned perpendicularly to the film plane$\langle S_{z}(T) \rangle$ are denoted very similarly and must not be confused.} 
The decisive quantity in Eq. (\ref{Gl1}) is the temperature dependent function
\begin{eqnarray}
C(T) = 1-\frac{1}{2S^2}\left( S(S+1) - \langle S^2_{z\prime}\rangle(T)\right).\label{Gl3}
\end{eqnarray}
By inspecting Eq.(\ref{Gl1}) it is obvious that $K_{2eff}(T)$  may change its sign as a function of temperature given that the temperature dependence of 
$C(T)$ is strong enough. Note again that no surface vs. interior competition is necessary to explain this change of sign of $K_{2eff}(T)$ which now turns 
out to be a consequence of the different temperature dependencies of the terms $\tilde{K_{2}}(T)$ and $\tilde{g_0}(T)$.
Thus a change of sign of $K_{2eff}(T)$  it also possible for translational invariant systems as will be  discussed in the next section.

To study the magnetic reorientation transition theoretically, one has to choose an appropriate model which describes the important physics of the systems under consideration. Furthermore certain approximations have to be tolerated and justified to solve the model in most cases.
It turns out that 
the (local-spin) Heisenberg model solved by the RPA approximation \cite{RPA}
describes very successfully magnetic properties
like spin wave spectra and Curie-temperatures of the transition metals Fe, Co
and Ni.
This finding is backed by ab-initio calculations with a Heisenberg model \cite{bruno1} and with a Gutzwiller wave function
approximation \cite{gutzwiller1,gutzwiller2,argument}.
In Ref. \cite{bruno1} the materials are described by an
Heisenberg
model. The real structure exchange parameters are calculated by use of the
magnetic force theorem \cite{PRL1,PRL2,PRL3}. The Heisenberg model is
solved
by the RPA approximation to calculate the Curie temperature as well as the
spin wave stiffness. Both quantities compare very well with the experimental ones.
In Refs. \cite{gutzwiller1,gutzwiller2,argument} the many body
problem posed by the strongly correlated d-electrons in the transition
metals is solved for T=0 using the LDA approximation to density functional theory and a
Gutzwiller wave function approach (see Refs. \cite{gutzwiller1} for the general method, Refs. \cite{gutzwiller1,gutzwiller2} for the application to Ni and Ref. \cite{argument} for the calculation of spin-waves in this scheme).
The properties of the transition metals are described very successfully by
this method (including Ni where DFT+LDA fails to reproduce some experimental
findings \cite{gutzwiller2}).
The results of this method can therefore be taken as a T=0 reference. It was
found that the spin-wave spectra compare very well to those of an
RPA-treatment of the Heisenberg model, but differ considerably to those obtained from a
mean-field like treatment of a band-model.

However, in such a theoretical description using the Heisenberg model the competition between the lattice and the  shape anisotropy will never lead to a change of 
sign of the effective anisotropy as long as  the lattice anisotropy is described by a second order term and the shape anisotropy is treated within mean-field (MF) theory \cite{magn2}, RPA \cite{RPA}, or thermodynamical perturbation 
theory \cite{nolt6}. Furthermore, it has been shown that a mean-field or RPA approximation is inappropriate \cite{jensen2000} for the local second order contribution to the lattice anisotropy which is given 
by $\sum_i K_2 S_{iz}^2$ (see e.g. Ref. \cite{wandlitz}).

In Refs. \cite{schwieger1,schwieger2,pini} a treatment of an extended Heisenberg model for film systems is proposed including an improved approximation 
for the lattice anisotropy  (generalized Anderson-Callen decoupling). It turns out that this theory yields quantitative agreement with numerically exact Quantum Monte Carlo calculations for the field induced 
reorientation transition of a monolayer with a positive and local second order contribution to the lattice anisotropy \cite{qm}. Furthermore  the decoupling of the local 
anisotropy term fulfills the exact limiting case for spin S=1/2, where 
$S_z^2 =1/4$ holds. Thus the commutators $[S_{+/-},S_z^2]$ trivially vanish and the anisotropy
does {\it not} influence the excitation spectrum for $S=1/2$ in this model. This behavior is not reproduced by an RPA or MF approximation.

The term $C(T)$ in Eq. \ref{Gl3} is a direct consequence of the special decoupling scheme for the local $K_2$-terms and  is missing if the lattice anisotropy term is treated in MF or RPA approximation.
The generalized Anderson-Callen theory for the lattice anisotropy is thus a decisive ingredient for the temperature dependent reorientation transition we discuss in this work.

\section{Theory}
In the following we want to present the theoretical approach for the film system:
As mentioned we want to concentrate on  translational invariant systems as e.g. a two-dimensional monolayer. Thereto the 
following Hamiltonian is used:
\begin{eqnarray}
H &=& -\sum_{ij}J_{ij}\mathbf{S}_{i}\mathbf{S}_{j} \label{hamilt}
 - \sum_{i}g_{J}\mu_{B}\mathbf{B}_0\mathbf{S}_{i}
 - \sum_{i}K_{2}S_{i z}^2+ \nonumber\\
 & & +\frac{1}{2}\sum_{ij}g_{0}\left(\frac{1}{r_{ij}^3}\mathbf{S}_{i}\mathbf{S}_{j} - \frac{3}{r_{ij}^5}(\mathbf{S}_{i}\mathbf{r}_{ij})(\mathbf{S}_{j}\mathbf{r}_{ij})\right)
\end{eqnarray}
This system is also a good model for films with finite thicknesses, but similar film parameters in all layers.

The first term describes the Heisenberg coupling $J_{ij}$ between magnetic spin moments $\mathbf{S}_{i}$ and $\mathbf{S}_{j}$ located 
at sites $i$ and $j$. Film thicknesses beyond monolayer can be absorbed into the parameters $J_{ij}$ which  can be used to realize the Curie temperature  of the film. The second term contains an external magnetic field $\mathbf{B}_0$ in arbitrary direction with the Land{$\rm\acute e$} factor (or more precisely the spectroscopic splitting factor) $g_{J}$ and the Bohr magneton $\mu_{B}$.  The third and fourth term represent second order lattice anisotropy and dipolar interaction, the latter leading to shape anisotropy. $S_{i z}$ is the $z$-component of $\mathbf{S}_{i}$ and is perpendicular to the film plane, $r_{ij}$ is the distance between lattice sites $i$ and $j$  in units of the nearest neighbor distance $a_0$.
The shape anisotropy favors in-plane orientation of the magnetization, the lattice anisotropy in-plane ($K_{2}<0$) or out-of-plane ($K_{2}>0$) 
orientation. Our Hamiltonian is similar to that used in Refs. \cite{wandlitz,jensen2000,schwieger1,schwieger2} for the investigation of the magnetic anisotropy and 
the field induced reorientation transition. 
This is the simplest Hamiltonian in which this new kind of the spin 
reorientation  occurs we discuss in this work. Higher order anisotropy terms as e.g. $K_4$-terms are not taken into account. It turns out that there are many film systems in which these higher 
order anisotropy terms can be neglected because $K_4 \ll K_2$ \cite{lindner}.

$g_0$ defines the dipolar coupling strength which is given for point-dipols by
\begin{eqnarray}
g^*_0 &=& (g_J\mu _B)^2/a_0^3\label{Gl2}.
\end{eqnarray}
For the nearest neighbor distance the value $a_0=1.81\text{\AA}$ is chosen which is the value of Fe grown on Cu(001).
 The Land$\rm\acute{e}$ factor $g_J$ is set $g_J=2.1$ \cite{eisengfactor}.
Note that in our model the spin quantum number $S$ is set to unity $S=1$.
The dipole coupling is caused by interaction of electrons with $S=1/2$ and the contribution
of the resulting shape-anisotropy to the effective anisotropy field scales with $\langle S_{z\prime}\rangle$ (see Eqs. \ref{Gl1} and \ref{g0effekt}).
Therefore the parameter $g^*_0$ has to be renormalized and one gets $g^*_0=3.44\rm\mu_BkG$.
Since the probability of finding the electrons and therefore the magnetic dipolar moment is not concentrated in one point $g^*_0$ may slightly be above this value. In the following the dipolar coupling strength is set to $g_0=3.8 {\rm \mu_BkG}$.
\\
To simplify calculations we consider nearest neighbor coupling only
\begin{eqnarray}
J_{ij} = \left\{ 
\begin{array}{cc} 
J & ~~(i),(j) \text{ n.n.}\\
0 & \text{elsewhere}
\end{array}\right.
\end{eqnarray}

The theory used here is a combination of RPA approximation for the nonlocal terms in Eq. (\ref{hamilt}) (Heisenberg exchange and dipolar interaction) and Anderson-Callen approximation for the local lattice anisotropy term.
For a detailed presentation  of this approach we  refer to Refs. \cite{schwieger1}, \cite{schwieger2} and \cite{pini}.
In the following we first comment on two improvements as compared to our earlier theoretical treatment and then summarize the main steps with the most important formulas. 
\begin{enumerate}
\item[a)] For convenience we neglected a certain type of Green functions in the works \cite{schwieger1,schwieger2} mentioned above, namely $\left\langle \left\langle S_{\mathbf{q}}^-,S_{-\mathbf{q}}^-\right\rangle \right\rangle$, $\left\langle \left\langle S_{\mathbf{q}}^+,S_{-\mathbf{q}}^+\right\rangle \right\rangle$, and $\left\langle \left\langle S_{\mathbf{q}}^-,S_{-\mathbf{q}}^+\right\rangle \right\rangle$. As was pointed out by Pini et al. in Ref. \cite{pini} these Green functions are needed for a more accurate description of easy-plane systems. Additionally, the spin wave excitations in the vicinity of the reorientation transition are described better for easy plane as well as for easy axis systems. 
\item[b)] Furthermore we improve  the treatment of the dipolar anisotropy. It was mean-field decoupled in Refs. \cite{schwieger1,schwieger2,pini} while all other terms were treated with the RPA \cite{RPA} (exchange and Zeeman terms) or the Anderson-Callen decoupling ($K_2$ terms) \cite{ac}.
Here we treat the dipolar coupling in complete analogy to the non-local exchange terms using the RPA decoupling. This gives a $\mathbf{q}$-dependent contribution  to the spin-wave energies. However, since the dipolar coupling is at least three orders of magnitudes smaller than the Heisenberg exchange coupling, this term can be neglected for all $\mathbf{q}>0$ where the Heisenberg exchange coupling $J$ determines the spin-wave energies. For the uniform mode $\mathbf{q}\rightarrow 0$, on the other hand, the influence of the Heisenberg exchange coupling vanishes and the dipolar term gets important, since it is of the same order of magnitude as the lattice anisotropy and the external field. Therefore we take into account only the $\mathbf{q}$=0 contribution of the dipolar coupling.
Note, that this treatment differs from a MF decoupling of the dipolar term. MF theory averages over the $q$-dependence. In our approximation the $q$-dependent dipolar contributions to the spin-wave energy are replaced by its $q=0$-component.

\end{enumerate}
Let us now shortly summarize the theory \cite{schwieger1,schwieger2} to solve the Hamiltonian (\ref{hamilt}).
\begin{enumerate}
\item[1.)] In general, the magnetization $M\propto \langle S_{z\prime}\rangle$ in the considered monolayer is not aligned parallel to the $z$-axis of our fixed coordinate system (the $z$-axis is chosen to be parallel to the film normal). First we rotate our coordinate system $\Sigma \rightarrow \Sigma \prime$ to align the $z\prime$-axis parallel to the magnetization. Note that due to the symmetry concerning the azimuthal angle $\phi$ we can always choose $\phi_M=\phi_{B_0}$.

\item[2.)] The polar angle of rotation $\theta$ is not fixed a priori. It is determined self-consistently. The condition for determining this angle is that the $z\prime$-contribution to the magnetization is approximately a constant of motion, i.e. after the approximations (decouplings of higher operator products) are performed
\begin{eqnarray}
\frac{d S_{z\prime}}{dt} &\stackrel{RPA+A.C.}{\longrightarrow}& 0
\end{eqnarray}
holds.
This condition leads to
\begin{eqnarray}
\left[S_{i z\prime},H\right]_- & \stackrel{RPA+A.C.}{\longrightarrow}& 0
\end{eqnarray}
and calculating the commutator and performing the decoupling procedures gives:

\begin{flalign}
g_{J}\mu_B(s_{\theta}B_{0z}-c_{\theta}B_{0x})+&\nonumber\\
+s_{\theta}c_{\theta}\langle S_{iz\prime}\rangle(2K_{2}& C -\frac{3}{2}g_{0}D) \stackrel{!}{=}0.\label{winkel1}
\end{flalign}

 The abbreviations
\begin{flalign*}
c_{\theta} = \cos\theta\\
s_{\theta} = \sin\theta\\
D = \frac{1}{N}\sum_{ij} \frac{1}{r_{ij}^3}
\end{flalign*}
are used.

\item[3.)] In the rotated system we write down the equations of motion for the following Green functions (GF):
\begin{eqnarray}
  \underline{G} &=& \left(\begin{array}{cc}
  G_{ij}^{+-} & G_{ij}^{--} \\
  G_{ij}^{++} & G_{ij}^{-+} \\
\end{array}\right)= \\
&=&
\left(\begin{array}{cc}
  \left\langle \left\langle S_{i}^{+\prime},S_{j}^{-\prime}\right\rangle \right\rangle & \left\langle \left\langle S_{i }^{-\prime},S_{j}^{-\prime}\right\rangle \right\rangle\\
  \left\langle \left\langle S_{i}^{+\prime},S_{j}^{+\prime}\right\rangle \right\rangle & \left\langle \left\langle S_{i }^{-\prime},S_{j }^{+\prime}\right\rangle \right\rangle\\
\end{array}\right)\label{GF}
\end{eqnarray}

In our earlier work \cite{schwieger1} only the GF $G_{ij}^{+-}$ was taken into account and not
the other GFs $G_{ij}^{-+}$, $G_{ij}^{--}$, $G_{ij}^{++}$ in the matrix (\ref{GF}). 
As mentioned above it was the  proposal of Pini et al. in Ref. \cite{pini} to take those GFs also 
into account.

To solve this system of equations higher operator products have to be decoupled. For non-local operator products, as e.g. $S_{iz\prime}S_{j-\prime}$, we use the RPA decoupling 
\begin{eqnarray}
S_{iz\prime}S_{j-\prime}\stackrel{RPA}{\longrightarrow}\langle S_{iz\prime}\rangle S_{j-\prime}+S_{iz\prime}\langle S_{j-\prime}\rangle
\end{eqnarray}
Since the magnetization is parallel to the $z\prime$-axis  
$\langle S_{ix\prime}\rangle = \langle S_{iy\prime}\rangle = \langle S_{i+\prime}\rangle = \langle S_{i-\prime}\rangle = 0$ holds in the above expression and the second summand vanishes. 

For local operator products ($K_2$ terms), the Anderson-Callen decoupling is used, which
leads in the primed system to
\begin{eqnarray}
S_{i  z\prime}S_{i  +\prime/-\prime} + S_{i  +\prime/-\prime}  S_{i  z\prime} \stackrel{A.C.}{\longrightarrow}&&\nonumber\\
 2 \langle S_{i  z\prime} \rangle C(T)&&S_{i  +\prime/-\prime}
\end{eqnarray}
with $C(T)$ as defined above.

As shown in Ref. \cite{schwieger1} only with this procedure the QMC results for the field induced reorientation of Ref. \cite{qm} can be quantitatively reproduced and the above-mentioned exact limiting case for S=1/2 is fulfilled. We checked, that these agreements still hold for the theory presented here, which is slightly modified as compared to Ref. [31].

\item[4.)] After decoupling and performing the two-dimensional Fourier transformation 
\begin{eqnarray*}
S_{i  +\prime;-\prime;z\prime} = \frac{1}{N}\sum_\mathbf{q} e^{-i \mathbf{q} \mathbf{R}_i}S_{\mathbf{q}  +\prime;-\prime;z\prime}
\end{eqnarray*}
we obtain the following system of equations:
\begin{equation}
\underline{G}_\mathbf{q} \left(E_\mathbf{q} \underline{I} - \underline{M}_\mathbf{q} \right) = \underline{X}_\mathbf{q}\label{matrix0}
\end{equation}
with
\begin{equation}
\underline{G}_\mathbf{q} = \left(\begin{array}{cc}
  G^{+-} & G^{++} \\
  G^{--} & G^{-+} \\
 \end{array}\right)_\mathbf{q}
 \end{equation}
\begin{equation}
\underline{M}_\mathbf{q} = \left(\begin{array}{cc}
  M^{+-} & M^{++}  \\
  M^{--} & M^{-+} \\
 \end{array}\right)_\mathbf{q}\label{matrix}
\end{equation}
\begin{equation}
    \underline{X}_\mathbf{q} = \left(\begin{array}{cc}
  2 \langle S_{z\prime}\rangle & 0  \\
  0 & -2 \langle S_{z\prime}\rangle \\
 \end{array}\right)_\mathbf{q}
\end{equation}

The matrix elements are given by
\begin{eqnarray}
M^{+-} &=& 2 J (p-\gamma_q) + g_{J}\mu_B(s_{\theta}B_{0x}+c_{\theta}B_{0z})+\nonumber\\
&&+(2K_{2} C(T) -\frac{3}{2}g_{0}D)(c_\theta^2-\frac{1}{2}s^2_\theta)\langle S_{ z\prime}\rangle\label{matrix1}\\
M^{--} &=&  - (K_{2}C(T) -\frac{3}{4}g_{0}D) s_{\theta}^2\langle S_{ z\prime}\rangle \label{matrix2}\\
M^{-+} &=& - \left(M^{+-}\right)\label{matrix3}\\
M^{++} &=& - \left(M^{--}\right)\label{matrix4}
\end{eqnarray}
where $p$ denotes the coordination number within the layer and $\gamma_q$ is a structural factor due to the two-dimensional Fourier transformation.

\item[5.)] 
Solving for the spin Green functions yields
weights $\chi_{\alpha}({\bf q})$ and excitation energies $E_{\alpha}({\bf q})=\hbar\omega_{\alpha}({\bf q})$  which in turn give the average magnon occupation  number
\begin{eqnarray}
	\varphi(T) = \frac{1}{N}\sum_q \frac{\chi_{+}(q)}{e^{\beta E_+(q)}-1}+\frac{\chi_{-}(q)}{e^{\beta E_-(q)}-1}\;.
\end{eqnarray}
The two terms describe the single-magnon excitations of the system for a given wave vector $q$, namely magnon creation ("+") and magnon annihilation  ("-").
 The magnetization (in the rotated frame) can then be computed from: \cite{magn2}
\begin{equation}    
\langle S_{z\prime}\rangle = \frac{(1+\varphi)^{2S+1}(S-\varphi)+\varphi^{2S+1}(S+1+\varphi)}{(1+\varphi)^{2S+1}-\varphi^{2S+1}}
\end{equation}
and 
\begin{eqnarray}
\langle S_{z\prime}^2\rangle &=& S(S+1)- \langle S_{ z\prime} \rangle (1+2\varphi)\label{szsquare}
\end{eqnarray}

\end{enumerate}

Eqs. (\ref{winkel1}) - (\ref{szsquare}) form a closed system of equations which can be solved self-consistently.

In Ref. \cite{jensen2} the authors also investigate the temperature dependence of the effective anisotropy using a similar Hamiltonian as (\ref{hamilt}) and a similar method: However, since the lattice anisotropy is chosen to be  large as compared to the shape anisotropy, no sign reversal of $K_{2eff}(T)$ is found.\footnote{We want to point out that their theory fails for an arbitrary direction of the external magnetic field. Some of our key results which we present in the next section are obtained for non-perpendicular alignment of $\mathbf{B}_0$ and film plane which requires the improved treatment of th $K_2$ anisotropy terms in the equation of motion for $\underline{G}_\mathbf{q}$. For a more detailed comparison of the two approaches we refer to Ref. \cite{schwieger1}.}

A generalization to multilayers for the description of more complex film systems is straightforward \cite{schwieger1}.
However, for our present purpose the monolayer Hamiltonian in Eq. (4) is the simplest Hamiltonian in which this new kind of spin reorientation occurs.

\section{Results}
\begin{figure}
\begin{center}
\resizebox{0.85\columnwidth}{!}{
\includegraphics{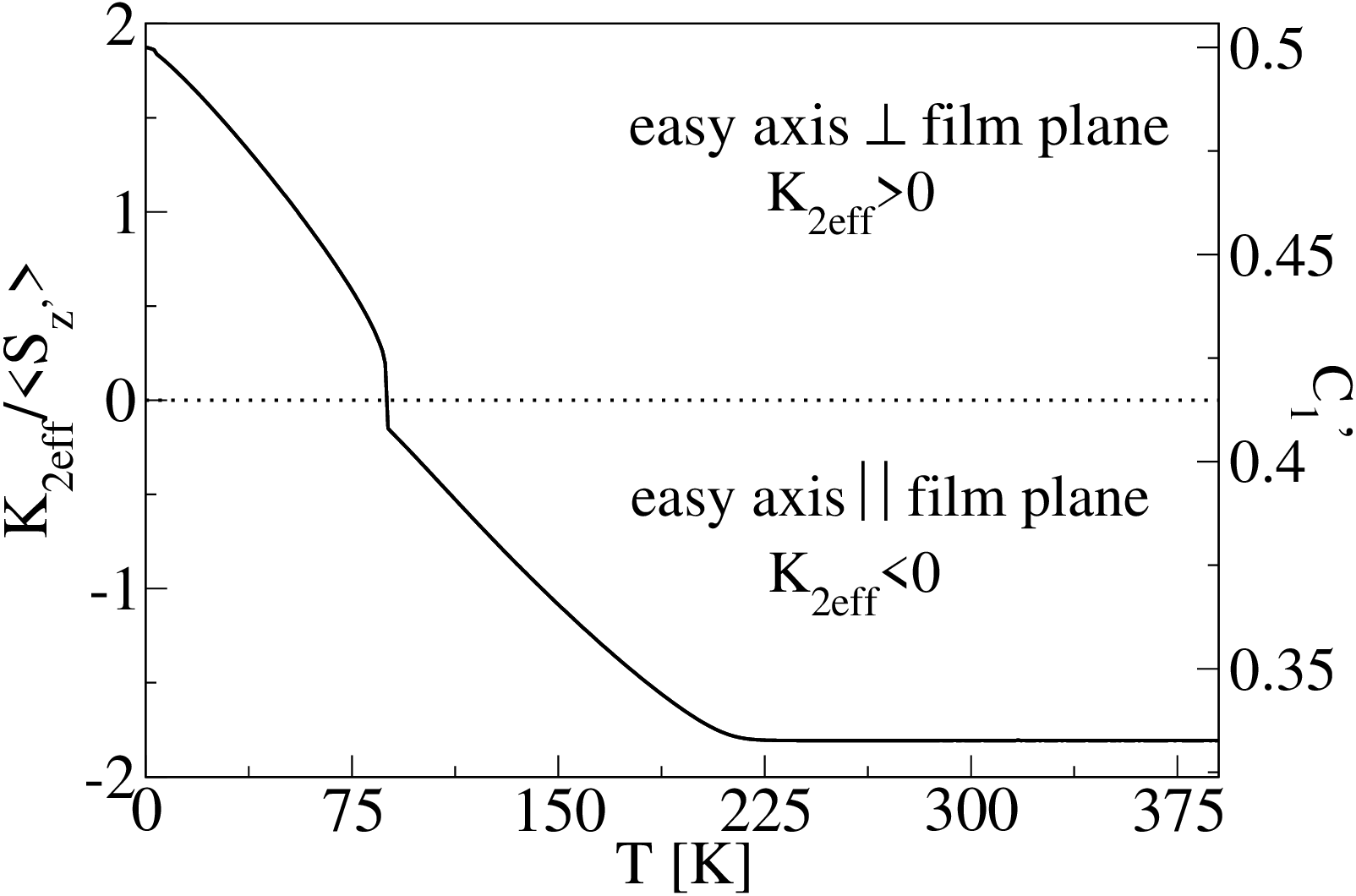}
}
\caption{ $K_{2eff}(T)/\langle S_{z\prime}\rangle(T)$ (left axis)  and $C(T)$ (right axis) are shown as functions of temperature. Both  decrease with increasing temperature. At the reorientation temperature $T_{reo}\approx 87K$ they have a sharp drop due to the reorientation and $K_{2eff}(T)$ changes its sign. Henceforward the easy magnetic axis of the monolayer is parallel to the film plane. }\label{figure1b}
\end{center}
\end{figure}

Let us now discuss the main results of our work. 
We choose the following parameters in our model study:
As mentioned before we set the spin quantum number $S=1$. The exchange parameter $J$ is chosen such that $T_c=200 \rm{K}$ holds \cite{eisentc} and the parameter $K_2$ is set to $11\mu_B\rm{kG}$ \cite{eisenk2}. 
Backed up by the point-dipole model the microscopic dipolar strength is  set to $g_0=3.8\mu_B\rm{kG}$. Therewith the chosen parameters are in a realistic range for e.g. thin Fe films.
In  Fig. \ref{figure1b}  the effective anisotropy field normalized by the  magnetization $K_{2eff}(T)/\langle S_{z\prime}\rangle(T)$ (left axis)  and $C(T)$ (right axis) are shown as functions of temperature. As seen in Eq. \ref{Gl1}, $K_{2eff}(T)/\langle S_{z\prime}\rangle$ depends linearly on $C(T)$ which depends on the temperature via the factor $\langle S^2_{z\prime}\rangle(T)$. The higher correlation function $\langle S^2_{z\prime}\rangle(T)$ depends itself on the magnetization  $\langle S_{z\prime}\rangle(T)$ and on the magnon number $\varphi$(T) (see Eq. \ref{szsquare}).
Due to the decrease of $C(T)$ the sign of the anisotropy field $K_{2eff}(T)$ changes at $T_{reo}\approx 87 K$ from positive to negative.

\begin{figure}
\begin{center}
\resizebox{0.85\columnwidth}{!}{
\includegraphics{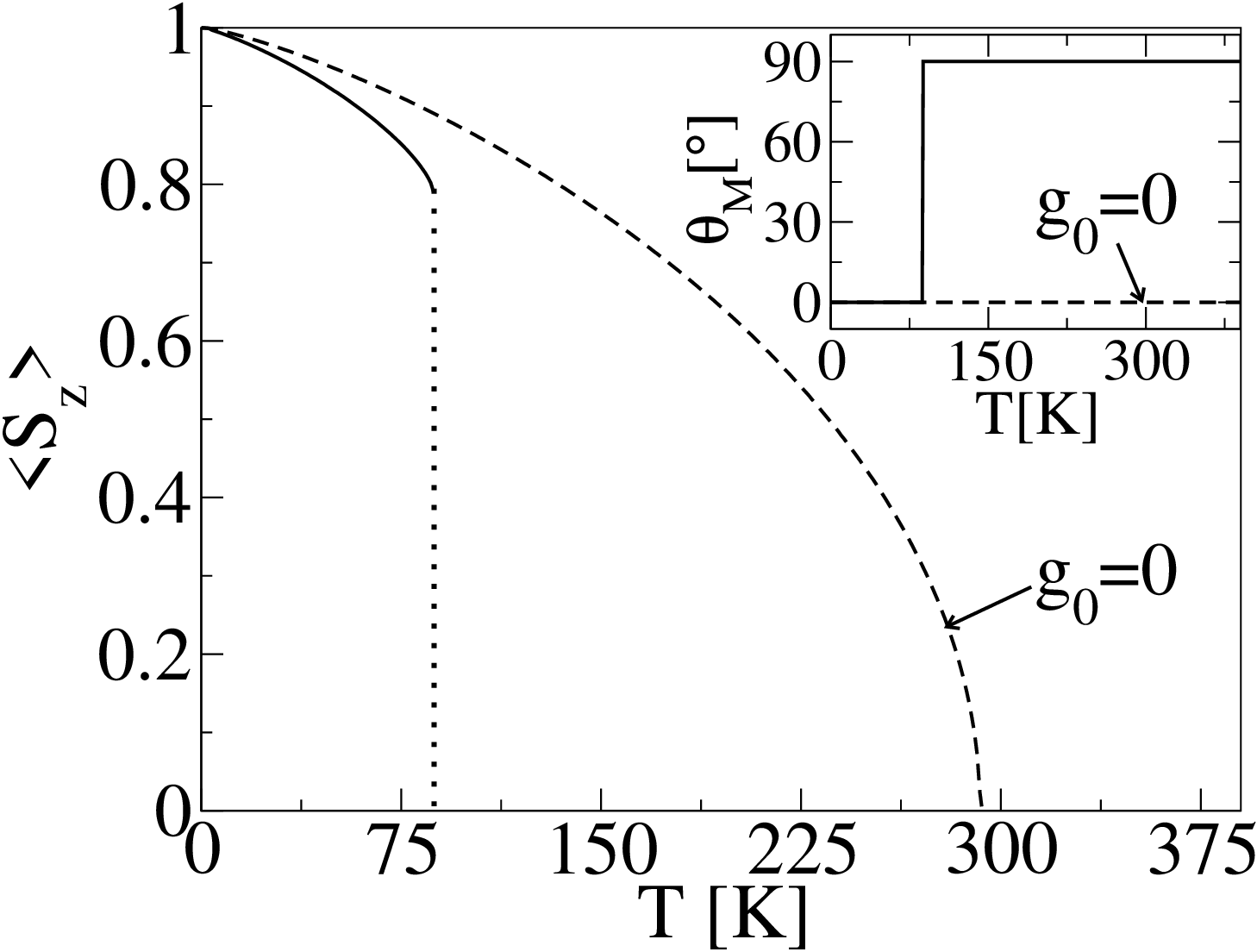}
}
\caption{ 
The z-component of the magnetization $\langle S_{z} \rangle (T)$ and the polar angle $\theta_M(T)$ (inset) are shown
 as  functions of temperature for $g_0=0$ (dashed line) as well as for $g_0=3.8\mu_B kG$ (solid line). }\label{figure2b}
\end{center}
\end{figure}

This sign reversal is accompanied by a sharp drop of $C(T)$ and $K_{2eff}(T)/\langle S_{z\prime}\rangle(T)$. The reason for this is a discontinuous drop of the magnetization at the reorientation due to the symmetry within the plane. 
At the point of the change of sign of $K_{2eff}(T)$, the magnetization sharply rotates from perpendicular to parallel alignment to the film plane as seen in Fig. \ref{figure2b}. Here the z-component of the magnetization $\langle S_{z} \rangle (T)$ as well as the polar angle $\theta_M(T)$ (inset) are shown  as  functions of temperature for $g_0=3.8\mu_B \rm{kG}$ (solid line). For comparison the $g_0=0$-line is also plotted (dashed line).

Strictly speaking for $T>T_{reo}$ the magnetization $\langle S_{z\prime}\rangle$ breaks down because according to the Mermin Wagner theorem a finite magnetization at finite $T$ is not consistent with gapless excitations, i.e. with $K_{2eff}=0$.
Therefore quantities $A$ like  $\langle S^2_{z\prime}\rangle$ and $K_{2eff}/\langle S_{z\prime}\rangle$ are calculated in the limit process
$A(B=0)=\lim_{Bx\to 0^+}A(B_x)$.

As seen in Fig. 2 the reorientation caused by the competition between shape and lattice anisotropy is missing for vanishing shape anisotropy $g_0=0$. Additionally the magnetization is reduced less with increasing temperature even before the reorientation transition $T<T_{reo}$. This is because in our case the lattice and shape contributions to the anisotropy field act against each other resulting in a higher anisotropy field for vanishing shape anisotropy.  Generally, the higher $K_{2eff}$ the weaker the magnetization is reduced with increasing temperature since the Curie temperature $T_c$ is a monotonically increasing function of the effective anisotropy strength.

\begin{figure}
\begin{center}
\resizebox{0.85\columnwidth}{!}{
\includegraphics{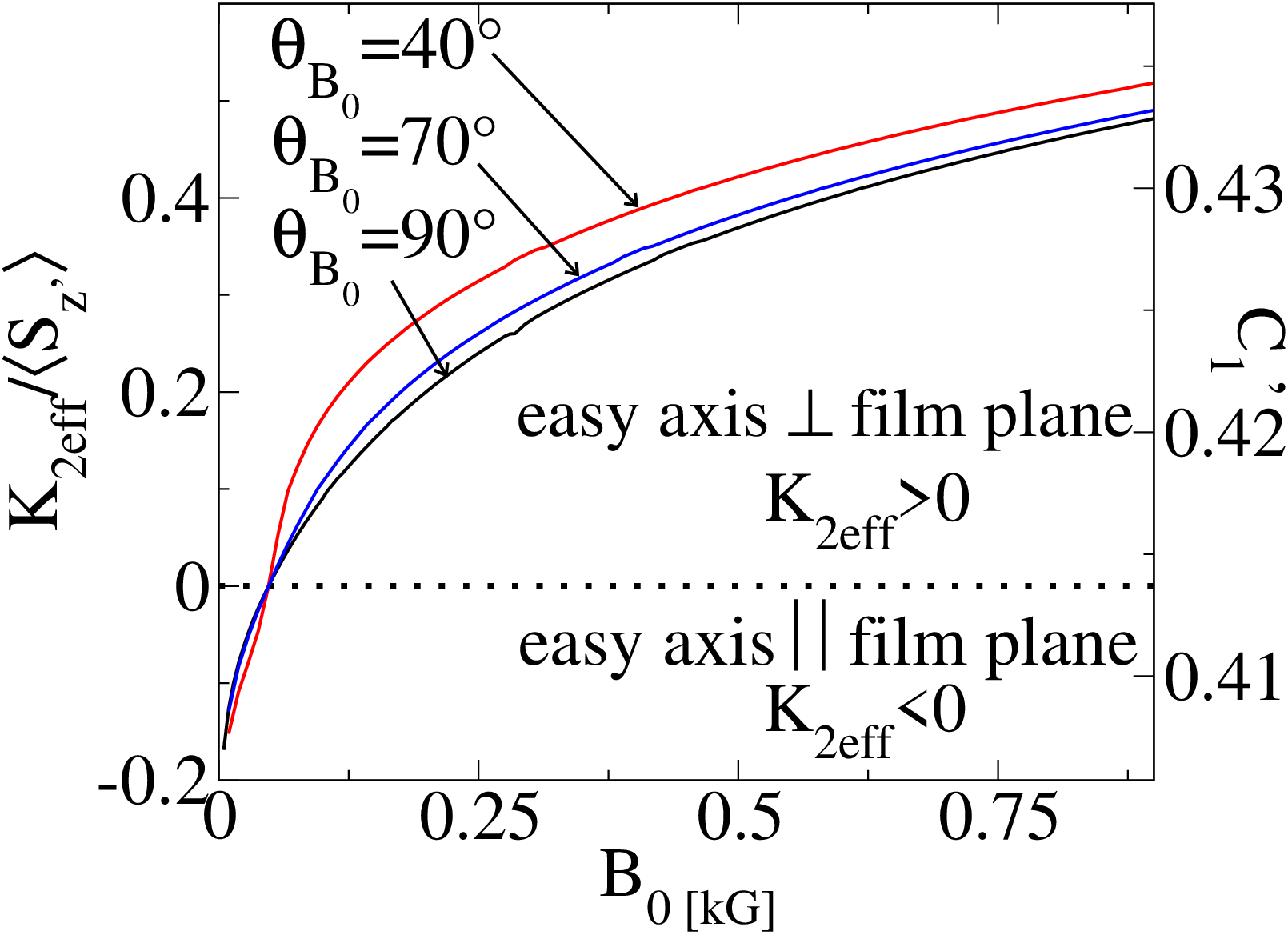}
}
\caption{$K_{2eff}(B)/\langle S_{z\prime} \rangle(B)$(left axis) and $C(B)$ (right axis) are shown for a fixed temperature $T_1=95 \rm{K} > T_{reo}$ for different directions of the external field. 
$K_{2eff}(B)/\langle S_{z\prime} \rangle(B)$ changes its sign from negative to positive caused by the external field.}\label{figure3b}
\end{center}
\end{figure}

In the following we want to discuss some important implications of this new type of temperature driven reorientation transition. 
Due to the fact that the magnetization $\langle S_{z\prime} \rangle$ and the magnon number $\varphi$ may also be very sensitive to changes of the external field, the quantities $C$ and therewith $K_{2eff}$ are also functions of the external field $B_0$.
This is demonstrated in Fig. \ref{figure3b} where $K_{2eff}(B)/\langle S_{z\prime} \rangle(B)$(left axis) and $C(B)$ (right axis) are shown for a fixed temperature $T_1=$95 $K > T_{reo}$. 
Both quantities increase with increasing external field. This is a direct consequence of the increasing norm of the magnetization $\langle S_{z\prime} \rangle$ and therewith the increase of its second moment $\langle S^2_{z\prime} \rangle$.
Due to the sensitive dependence of $K_{2eff}(B)$ on $C(B)$ it leads to a change of sign of $K_{2eff}(B)$ from negative to positive caused by the external field. Therefore the temperature driven change of sign of the effective anisotropy as discussed above may be compensated by an external field aligned in \textit{arbitrary} direction.
\\
 The two discussed effects could lead to the following scenario: At a low temperature $C(T<T_{reo})$ is large enough for $K_{2eff}$ to be positive. $C(T)$ then decreases with temperature until at $T_1>T_{reo}$ $K_{2eff}$ gets negative. 

\begin{figure}
\begin{center}
\resizebox{0.85\columnwidth}{!}{
\includegraphics{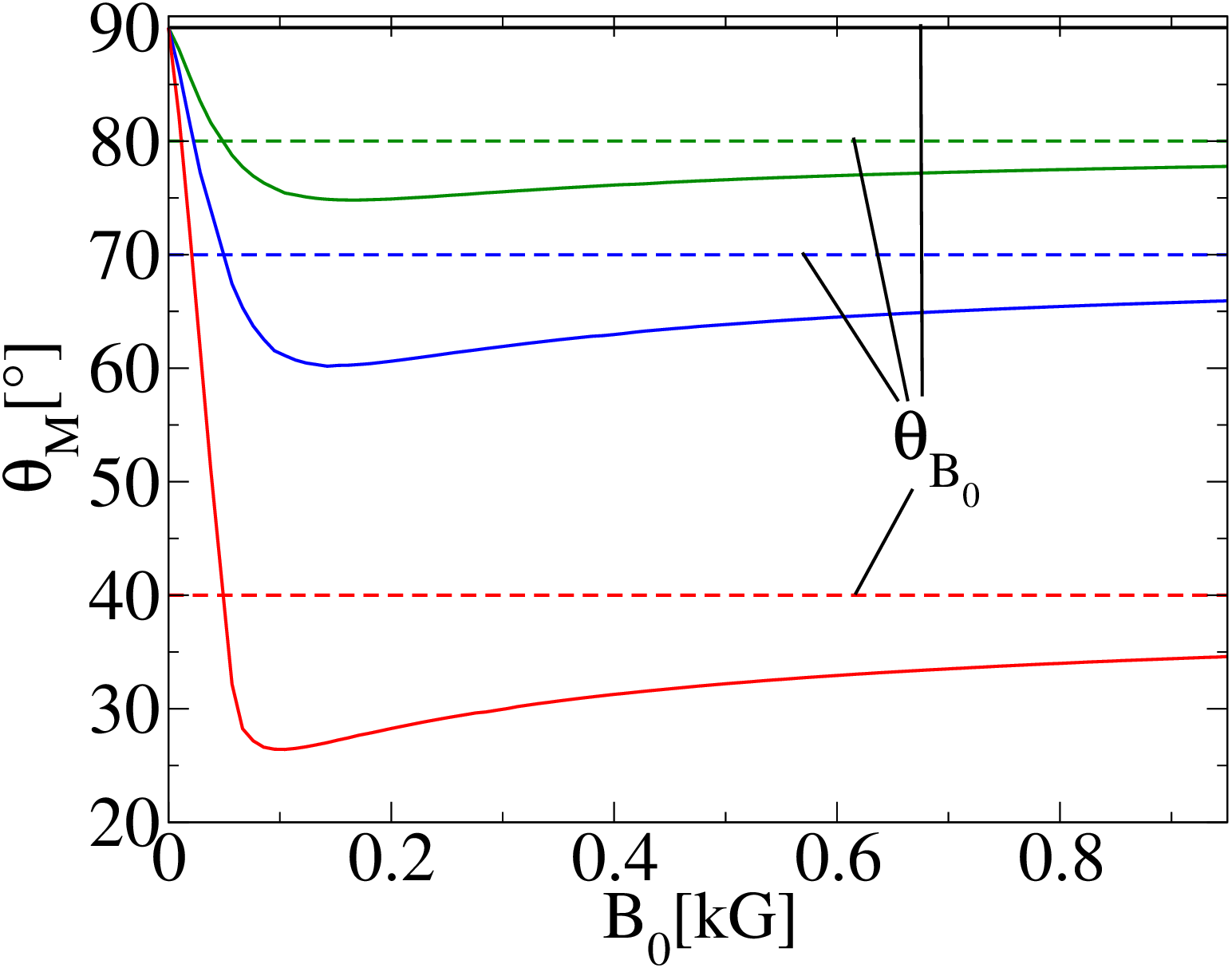}
}
\caption{The polar angle of the magnetization $\theta_M(B)$ is shown as a function of the external field (filled lines) for different directions of the external fields (dashed lines) for a fixed temperature $T_1=95\rm{K}>T_{reo}$. At the crossing point the effective anisotropy changes its sign from negative (easy axis $\parallel$ film plane) to positive (easy axis $\perp$ film plane).}\label{figure4b}
\end{center}
\end{figure}

  Now, if the temperature $T$ is kept fix to $T_1=95 \rm{K}$ an external field applied in any direction can lead to an increase of the norm of the magnetization $\langle S_{z\prime}\rangle(B)$, its second moment $\langle S^2_{z\prime}\rangle(B)$ and therewith $C(B)$. 
Thus $K_{2eff}$ can get positive again for $B>B_{reo}(T)$.
Having these considerations in mind one can understand the remarkable curves in Fig. \ref{figure4b}. An applied external field causes the magnetization to rotate from an alignment parallel to the film plane far beyond the direction of the external field, before magnetization and external field become parallel for large $B_0$.
For small fields the effective anisotropy is still negative (see Fig. 3) and therefore the easy magnetic axis is in-plane. Therefore the magnetization is aligned between the magnetic easy axis and the direction of the external field ($\theta_{B_0}<\theta_M<90\grad$). For a specific field $B_{crit}$ the effective anisotropy vanishes (see Fig. 3) and the magnetization is aligned parallel to the magnetic field (crossing points of solid and dashed lines in Fig. 4). Then for increasing magnetic field the effective anisotropy becomes positive and the \textit{new} easy magnetic axis is out-of-plane.
Hence the magnetization is aligned between the new easy magnetic axis and the direction of the external field ($0\grad<\theta_M<\theta_{B_0}$). For $B\approx B_{crit}$ one thus obtains a curious result: The magnetization is aligned parallel to the external field at the critical field and rotates {\it away} from the external field direction when the external field is further enhanced. It appears as if the magnetization was repelled by the external magnetic field.

\section{Summary and Conclusions}
We have proposed a new type of temperature induced reorientation transition  caused by competing lattice and shape anisotropy. It differs  from the well-known reorientation transitions caused by competing surface and bulk contributions and can occur also in translational invariant systems or uniform film systems (i.e. where the lattice anisotropies are similar in all layers).
 We use a theory which starts from an extended Heisenberg model including the exchange coupling, the Zeeman term, the second order lattice anisotropies, and the magnetic dipole coupling. A decoupling scheme for the local $K_2$-terms is used which was shown to yield very good agreement with QMC calculations before \cite{schwieger1}. Obtaining different temperature dependencies for the lattice and the shape contribution to the effective anisotropy,  we find a change of sign of the effective anisotropy field $K_{2eff}(T)$ with increasing temperature.

Let us comment on the relevance on the proposed mechanism for experiments:
First, using a realistic magnitude of the dipolar  interaction ($g_0=3.8\mu_B$kG), the lattice anisotropy must be approximately $K_2\sim10\mu_B$kG (for $S=1$) to obtain a temperature dependent magnetic reorientation transition. This value is a realistic magnitude for e.g. thin  Fe/Cu(001) films \cite{eisenk2}. Therefore 
our findings might be relevant for real materials. 
Secondly, note that this kind of transition is superimposed to the reorientation due to surface-bulk competition and  possibly complicates the interpretation of experiments. 
\\
A very important consequence might follow for experiments which measure the effective anisotropy $K_{2eff}(T,B)$ using external magnetic fields (e.g. MOKE, FMR). As we point out $K_{2eff}(T,B)$ may depend sensitively on the external field in the vicinity of the discussed reorientation transition and may have another sign as the $B=0$-effective anisotropy $K_{2eff}(T,0)$. It would be interesting to investigate experimentally this effect by  measuring $K_{2eff}(B)$, e.g. using FMR with different microwave frequencies. 

\begin{acknowledgement}
Productive discussions with K. Baberschke, K. Lenz and P.J. Jensen are gratefully acknowledged. 
\end{acknowledgement}

\end{document}